\documentstyle[11pt,aaspp4]{article} 
\begin{document}
\title{On the Graviton Mass} 
\author{Andrei Gruzinov}
\affil{Physics Department, New York University, 4 Washington Place, New York, NY 10003}

\begin{abstract}

It was suggested that observations of the solar system exclude massive gravity, in the sense that the graviton mass must be rigorously zero. This is because there is a discontinuity in the linearized gravity theory at graviton mass equal to zero. The linearized Schwarzschild metric is not recovered for infinitesimal graviton mass, contradicting observations on light deviation by the Sun and Mercury perihelion advance. It was then argued that non-perturbative effects make the massive gravity theory continuous in the graviton mass. Both the original suggestion and its refutation were based on a non-covariant and linearized action, and the physical interpretation of these results remained questionable. Here we use a covariant quasi-massive gravity theory that is known to be discontinuous in the graviton mass in the linear approximation. We show that non-perturbative effects do restore the continuity; the weak-field Schwarzschild solution is recovered in the limit of small graviton mass. We also show that weak-field Schwarzschild with matter is recovered for infinitesimal graviton mass. Thus: Observations of the solar system only put an upper limit on the graviton mass (in the context of the gravity theory that we use, inverse graviton mass, as measured at distances of order inverse graviton mass, is $\gtrsim 100$ Mpc). But graviton can be massive, with a cosmologically interesting mass.

\end{abstract}

\section{What's wrong with a non-zero graviton mass?}

Zakharov (1970) and van Dam \& Veltman (1970) (ZvDV) suggested that observations of the solar system tell us that the graviton mass must be rigorously zero. One would usually expect that observations can give just an upper limit, but in the linearized gravity the ZvDV theorem is indeed correct as we explain below using a non-covariant linearized massive gravity. In \S2 a covariant version of the ZvDV theorem is given in the quasi-massive (induced) gravity of Dvali, Gabadadze \& Porrati (2000). 

Consider linearized massive gravity on the flat background: the metric tensor is $g_{ik}+h_{ik}$, $g_{ik}={\rm diag}(1,-1,-1,-1)$, $|h_{ik}|\ll 1$. The Lagrangian should be of the form
\begin{equation}
{\cal L} = {\cal L}_E-{1\over 4}m^2(h_{ik}h^{ik}+ah^2).
\end{equation}
Here ${\cal L}_E$ is the quadratic part of the Einstein Lagrangian and $h=h_i^i$. Energy is positive if and only if $a=-1$, which is the Pauli-Fierz mass term. But then, the graviton propagator is 
\begin{equation}
P_{iknm}={{1\over 2}g_{in}g_{km}+{1\over 2}g_{im}g_{kn}-{1\over 3}g_{ik}g_{nm}\over k^2-m^2},
\end{equation}
while the standard massless propagator is
\begin{equation}
P_{iknm}={{1\over 2}g_{in}g_{km}+{1\over 2}g_{im}g_{kn}-{1\over 2}g_{ik}g_{nm}\over k^2}.
\end{equation}
The two propagators are different by order one even in the limit $m=+0$ --- the ZvDV theorem. 

In the solar system, gravity is weak, and it appears that the use of the linearized approximation is justified. But then observations (light deviation and Mercury perihelion advance) rule out massive gravity. The graviton mass should be rigorously zero. That observations can measure the graviton mass to an absolute accuracy is strange, and indeed Vainshtein (1972) showed that non-linear corrections restore the continuity in the graviton mass. 

Both the ZvDV theorem, and its non-perturbative refutation are based on a non-covariant theory. The physical interpretation of these results in the GR framework is unclear. One could argue that the absence of a covariant massive gravity theory indicated that $m=0$. Also, Vainshtein (1972) had to calculate non-linear corrections using a partly linearized action -- a questionable procedure. In our view, the problem of the graviton mass remained open.

Recently, Dvali, Gabadadze \& Porrati (2000) introduced the first covariant theory of (quasi) massive gravity, and showed that the ZvDV theorem holds in their theory too (\S 2). Nevertheless, in \S 3 we show that this massive gravity theory is consistent with the solar system observations if the graviton is light enough. We show that corrections to the Schwarzschild metric of the Sun vanish for very light gravity (a similar result holds for a string,  Lue (2001)). Interestingly, the resulting limit on the graviton mass does not rule out possible cosmological effects of this massive gravity (Deffayet et al, 2001). 

\section{ZvDV for Schwarzschild metric in induced gravity}

The action of induced gravity is (Dvali, Gabadadze \& Porrati 2000):
\begin{equation}
-\int d^4x \sqrt{-g}R~-~L^{-1}\int d^5x \sqrt{-{\rm  g}}{\cal R}.
\end{equation}
Here the first term is the Einstein action, $L$ is a constant with dimension of length, the second term is a 5D Einstein action, and $g_{ik}={\rm g}_{ik}$ at $x^4=0$, $i,k=0...3$.  This action is covariant, that is covariant in the 4D space-time $x^4=0$, and this is all we need. For our purposes, reality of the fifth dimension is irrelevant. We will be interested only in the resulting 4D geometry of $x^4=0$. The last term is then analogous to the mass-term, with $m^2\sim (Lr)^{-1}$. 

The empty-space Einstein equation becomes
\begin{equation}
R^{ik}-{1\over 2}Rg^{ik}~+~{2\sqrt{-{\rm  g}}\over L\sqrt{-g}}\{~ {\cal R}^{ik}-{1\over 2}{\cal R}{\rm g}^{ik} ~\}=0,
\end{equation}
where the last term should be understood as follows. The Jacobian $\sqrt{-{\rm  g}}$ is calculated at $x^4=0$. The Einstein tensor ${\cal R}^{ik}-{1\over 2}{\cal R}{\rm g}^{ik}$ is calculated at $x^4=+0$, and the operation $\{ ... \}$ leaves only the $\delta$-function part of the 5D piece. This operation is defined as follows: $\{ a\partial _4\partial _4A+b\partial _4B+C \}~=~a\partial _4A$, where expressions $a$, $A$, $b$, $B$, and $C$ do not contain derivatives over $x^4$. For $x^4>0$, 
\begin{equation}
{\cal R}_{IK}=0, 
\end{equation}
where $I,K=0...4$. ${\rm g}_{IK}$ are symmetric functions of $x^4$. 

Consider modifications of the Schwarzschild metric in induced gravity. Take the line element of the form
\begin{equation}
ds^2={\rm e}^{\nu }dt^2-{\rm e}^{\lambda }dr^2-{\rm e}^{\mu }(\sin ^2\chi d\theta ^2+\sin ^2\chi \sin ^2\theta d\phi ^2+d\chi ^2),
\end{equation}
with $x^4=r\cos \chi$. The three unknowns $\nu$, $\lambda$, and $\mu$ are functions of $r$ and $\chi$.

The nontrivial Ricci components are 
\begin{equation}
2{\cal R}^0_0={\rm e}^{-\lambda }\left( \nu '' +{1\over 2}\nu '(\nu '-\lambda '+3\mu ')\right) +{\rm e}^{-\mu }\left( \ddot{\nu }+{1\over 2}\dot{\nu }(\dot{\nu }+\dot{\lambda }+\dot{\mu }+4\cot \chi )\right) ,
\end{equation}
\begin{equation}
2{\cal R}^1_1={\rm e}^{-\lambda }\left( \nu '' +3\mu ''+{1\over 2}\nu '(\nu '-\lambda ')+{3\over 2}\mu '(\mu '-\lambda ')\right) +{\rm e}^{-\mu }\left( \ddot{\lambda }+{1\over 2}\dot{\lambda }(\dot{\nu }+\dot{\lambda }+\dot{\mu }+4\cot \chi )\right) ,
\end{equation}
\begin{equation}
2{\cal R}^2_2=2{\cal R}^3_3={\rm e}^{-\lambda }\left( \mu '' +{1\over 2}\mu '(\nu '-\lambda '+3\mu ')\right) +{\rm e}^{-\mu }\left( \ddot{\mu }+{1\over 2}\dot{\mu }(\dot{\nu }+\dot{\lambda }+\dot{\mu })+\cot \chi (\dot{\nu }+\dot{\lambda }+3\dot{\mu })-4\right) ,
\end{equation}
\begin{equation}
2{\cal R}^4_4={\rm e}^{-\lambda }\left( \mu '' +{1\over 2}\mu '(\nu '-\lambda '+3\mu ')\right) +{\rm e}^{-\mu }\left( 2\ddot{\mu }+\ddot{\nu }+\ddot{\lambda }+{1\over 2}\dot{\nu }(\dot{\nu }-\dot{\mu })+{1\over 2}\dot{\lambda }(\dot{\lambda }-\dot{\mu })+2\cot \chi \dot{\mu }-4\right) ,
\end{equation}
\begin{equation}
2{\cal R}_{14}=-\dot{\nu }'-2\dot{\mu }'+\dot{\lambda }({1\over 2}\nu '+\mu ')+{1\over 2}\dot{\nu }(\mu '-\nu '),
\end{equation}
where prime and dot are derivatives over $r$ and $\chi$.

The 4D Ricci tensor is 
\begin{equation}
2R^0_0={\rm e}^{-\lambda }\left( \nu '' +{1\over 2}\nu '(\nu '-\lambda '+2\mu ')\right) ,
\end{equation}
\begin{equation}
2R^1_1={\rm e}^{-\lambda }\left( \nu '' +2\mu ''+{1\over 2}\nu '(\nu '-\lambda ')+\mu '(\mu '-\lambda ')\right) ,
\end{equation}
\begin{equation}
2R^2_2=2R^3_3={\rm e}^{-\lambda }\left( \mu '' +{1\over 2}\mu '(\nu '-\lambda '+2\mu ')\right) -2{\rm e}^{-\mu }.
\end{equation}

In the linear approximation, we take $\mu \rightarrow 2\ln r+\mu$ and linearize in $\nu$, $\lambda$, and $\mu$. Then equations (6) are solved by
\begin{equation}
\nabla_{(5)}^2\nu\equiv \nu ''+{3\over r}\nu '+{1\over r^2}(\ddot{\nu }+2\cot \chi \dot{\nu })=0,
\end{equation}
\begin{equation}
\mu =a+b\cos \chi -{\nu \over 2},
\end{equation}
\begin{equation}
\lambda =ra'+a+rb'\cos \chi -{\nu \over 2},
\end{equation}
where $a(r)$ and $b(r)$ are arbitrary functions of $r$.

Write the linearized equations (5) in the form 
\begin{equation}
2R^0_0+{2r\over L}\{ 2{\cal R}_0^0+{\cal R}_4^4\} =0,
\end{equation}
and similar equations for 11 and 22 components. We get, at $\chi =\pi /2$, 
\begin{equation}
\nu ''+{2\over r}\nu '-{3\over 2Lr}\dot{\nu } +{1\over L}(b'+{2b\over r})=0,
\end{equation}
\begin{equation}
{1\over r}\nu '-{3\over 2Lr}\dot{\nu } -{1\over L}(3b'+{2b\over r})=0,
\end{equation}
\begin{equation}
\nu ''+{1\over r}\nu '-{3\over Lr}\dot{\nu } -{2\over L}(b'+{4b\over r})=0.
\end{equation}
These equations are solved by 
\begin{equation}
\nu ''+{2\over r}\nu '-{2\over Lr}\dot{\nu }=0,
\end{equation}
\begin{equation}
b=-{L\over 4}\nu '.
\end{equation}

Now consider the small graviton mass limit, which corresponds to large $L$ in induced gravity. Equations (16), (23) give, for $\chi \leq \pi /2$:
\begin{equation}
\nu = -{2\chi\over \pi \sin \chi}{r_g\over r},
\end{equation}
where $r_g$ is a constant. Then (24) gives $b=-{Lr_g\over 4r^2}$. Taking $a=-{r_g\over 2r}$, so as to make $\mu =0$ in the 4D space-time, we get $\lambda ={r_g\over 2r}$ in 4D. This is to be compared with the linearized Schwarzschild's $\lambda ={r_g\over r}$. Thus ZvDV holds in induced gravity. 

But: is the linearization justified? At the very least, all metric components should be much less than unity. This is violated by the linearized solution. Indeed, for large $L$, the full 5D solution (given by (17), (18), (24)) is, for $\chi \leq \pi /2$:
\begin{equation}
\lambda = {\chi\over \pi \sin \chi}{r_g\over r}+\cos \chi {Lr_g\over 2r^2},
\end{equation}
\begin{equation}
\mu = \left( {\chi\over \pi \sin \chi}-{1\over 2}\right) {r_g\over r}-\cos \chi {Lr_g\over 4r^2}.
\end{equation}
In 5D, $\mu$ and $\lambda$ are large for large $L$. Even though we are only interested in 4D, the linearized theory is not applicable at large $L$. What happens at large L is explained in \S 3. It turns out that the linear theory works only for $b^2\ll a$, that is for $L^2r_g\ll r^3$.  

\section{Schwarzschild solution at large L}
To analyze the large $L$ limit of the weak ($r\gg r_g$) gravity, the full non-linear theory is not needed. It is sufficient to check that the following substitution
\begin{equation}
\mu = 2\ln r+b\cos \chi +{1\over 8}b^2\cos 2\chi +a -{\nu \over 2},
\end{equation}
\begin{equation}
\lambda = rb'\cos \chi +{1\over 8}rb'(2b-rb')\cos 2\chi +ra'+a+{1\over 8}(b^2-r^2b'^2) -{\nu \over 2},
\end{equation}
where $a$, $b$ are arbitrary functions of $r$, and $\nu $ satisfies 
\begin{equation}
\nabla_{(5)}^2\nu=0,
\end{equation}
solves ${\cal R}_{IK}=0$ up to $O(b^2)=O(a)=O(\nu)$. 

Now choosing $\nu$ given by (25), and $b=\sqrt{2r_g/r}$, $a=-r_g/(4r)$, we recover the weak-field Schwarzchild in the 4D space-time. And thus we also solve the $L=\infty$ limit of equations (5). 

The weak-field Schwarzschild with matter is also recovered in the large $L$ limit of induced gravity, if the proper energy density $\epsilon$ is positive. The weak-field Schwarzschild with matter is $\mu =2\ln r$, $r\nu '=\lambda$, $r\lambda =\int dr~r^2\epsilon$. In induced gravity we choose $a$ so as to give $\mu =2\ln r$. Then $\lambda =r\nu '/2+b^2/4$. Thus, the $L=\infty$ limit of induced gravity recovers the weak-field Schwarzschild with matter if $\lambda \geq r\nu '/2$, or $\lambda \geq 0$, which is true for $\epsilon\geq 0$. 

The finite $L$ corrections to the Schwarzschild solution are given by (5) (that is by (20)): 
\begin{equation}
{\delta \nu \over \nu }~=~ -\left( {8r^3\over L^2r_g}\right) ^{1/2}.
\end{equation}
For the solar system, assuming that fractional corrections at about $5$AU are $\lesssim 10^{-8}$ (Talmadge et al 1988), we get $L\gtrsim 100$Mpc. But gravity can be massive.

\acknowledgements

I thank my NYU colleagues:  Gia Dvali for discussions that initiated this study, Cedric Deffayet, and Arthur Lue for useful discussions, and Massimo Porrati for advice on how to derive the Pauli-Fierz mass term (use rest frame).

\end{document}